\documentclass[prb,twocolumn,aps,superscriptaddress]{revtex4-2}
\usepackage{comment}
\usepackage{graphicx}
\usepackage{dcolumn}
\usepackage{bm}
\usepackage{amssymb}
\usepackage{amsmath}
\usepackage{physics}
\usepackage{sublabel}
\usepackage[utf8]{inputenc}
\usepackage{hyperref}
\usepackage[usenames, dvipsnames]{color}
\usepackage[english]{babel}
\usepackage[T1]{fontenc}
\usepackage{bbm}

\usepackage{float}
\usepackage{verbatim}
\hypersetup{colorlinks = true, linkcolor=blue, citecolor=blue, urlcolor=blue}

  \makeatletter
    \renewcommand\@make@capt@title[2]{%
     \@ifx@empty\float@link{\@firstofone}{\expandafter\href\expandafter{\float@link}}%
      {\textsc{#1}}\@caption@fignum@sep#2\quad}%
    \makeatother
\begin{document}

\preprint{APS/123-QED}
\title{Effectuating tunable valley selection via multi-terminal monolayer graphene devices}
%\title{Tuning valley selection using multi-terminal graphene nanoribbon device structures}
%\title{Tunable multi-terminal graphene nanoribbon valley-selective devices}

\author{Shrushti Tapar}
\author{Bhaskaran Muralidharan}%
 \email{bm@ee.iitb.ac.in}
\affiliation{Department of Electrical Engineering, Indian Institute of Technology Bombay, Powai, Mumbai-400076, India}
%\date{\today}% It is always \today, today, %  but any date may be explicitly specified

\begin{abstract}
Valleytronics using two-dimensional materials opens unprecedented opportunities for information processing with the valley polarizer being a basic building block. Paradigms such as strain engineering, the inclusion of line defects, and the application of electrostatic-magnetic fields extensively explored for creating valley polarization suffer from limitations like smaller transmission or the lack of polarization directionality. We propose an all-electrical valley polarizer using zigzag edge graphene nanoribbons in a multi-terminal device geometry, that can be gate-tuned to operate along two independent regimes: (i) terminal-specific valley filter that utilizes bandstructure engineering, and (ii) parity-specific valley filter that exploits the parity selection rule in zigzag edge graphene. We show that the device exhibits intriguing physics in the multimode regime of operation that affects the valley polarization and hence investigate various factors affecting the polarization in wide device geometries, such as, optical analogs of graphene Dirac fermions, angle-selective transmission via p-n junctions, and the localization of edge states. We optimize the geometry of the proposed device to achieve maximum valley polarization, thereby, paving the way toward a physics based tunable valleytronic device design using monolayer graphene.
\end{abstract}


%\keywords{Suggested keywords}%Use the 22show keys class option if the %display desired
\maketitle
 
%\tableofcontents

\section{\label{sec:level1}Introduction}
\indent Valleytronics- the ability to control and manipulate the valley degree of freedom opens new possibilities for energy-efficient devices for both classical and quantum information processing \cite{vitale2018valleytronics,ang2017valleytronics}. Two-dimensional (2D) materials with momentum-separated inequivalent valleys, such as transition metal di-chalcogenides (TMDs) \cite{schaibley2016valleytronics}  and graphene \cite{schirber2021valley,qiao2014quantum}, are at the forefront of current research on valleytronic applications. To access the valleys \cite{zhao2021valley}, many techniques such as introducing circularly polarized light\cite{xia2017valley,friedlan2021valley,caruso2022chirality,mccreary2017understanding}, external magnetic fields \cite{gamayun2018valley}, defect engineering \cite{refaely2018defect,liu2013controllable,gunlycke2011graphene}, and strain engineering\cite{wang2020strain,chauwin2022strain,wu2011valley,milovanovic2016strain,zhao2020enhanced} are typically employed. However, in light of the current technological compatibility, an all-electrical control of the valley degree of freedom is typically desired. In 2D materials, valleys are electrically accessible via the presence of uneven Berry curvatures in the vicinity of the $K$ and $K’$ points \cite{hsu2015optically,feng2019engineering,zhao2017enhanced,ye2016electrical}. While broken inversion symmetry in TMDs naturally induces a finite bandgap and a valley-contrasting Berry curvature across the valleys \cite{zhou2019spin}, it is monolayer graphene that is a highly pursued and mature system for electronics applications \cite{yu2017analyzing,bolotin2008ultrahigh}. Owing to this, it is certainly worth devling into the physics of harnessing the valley degree of freedom in monolayer graphene.\\
\indent The presence of an inversion symmetry in monolayer graphene, however, complicates its potential for valleytronic applications. Given that the building block for valleytronics is the valley polarizer, several methods have been proposed for creating a high degree of polarization using monolayer graphene. These include sublattice staggered potentials \cite{zhou2007substrate} that break the inversion symmetry \cite{xiao2007valley,qiao2011spin}, line-defect engineering \cite{liu2013controllable,gunlycke2011graphene}, the application of uniform strain \cite{chauwin2022strain,wu2011valley} and magnetic fields \cite{yesilyurt2016perfect,zhai2010magnetic,zhai2012valley,lu2016valley} and the inclusion of various strain profiles \cite{milovanovic2016strain,cavalcante2016all,zhai2018local,li2020valley,wu2018quantum,hsu2020nanoscale,low2010strain}, to name a few. Utilizing electrostatic potential barriers \cite{wang2017valley,asmar2017minimal} that exploit the anisotropy of the Fermi surfaces for valley splitting caused by trigonal warping \cite{pereira2008valley,garcia2008fully} is also a crucial aspect. While the aforementioned approaches exhibit polarization, they are limited by  directionality and low transmission, and are hard to realize on-chip. The object of this paper is hence to propose an all-electrical valley polarizer based on monolayer graphene nanoribbons (GNR) that can be further be gate-tuned to operate in two regimes; Regime-1: Terminal specific valley filter (TSVF) attributed to band structure engineering, and Regime-2: Parity specific valley filter (PSVF) which uses the parity selection rule to operate as a valley filter. \\
\indent  Using the concept of the valley valve and the valley filter effects in a single-moded setup \cite{rycerz2007valley}, it was later also established as to how the parity effect, i.e., the number of zig-zag rows plays a critical role \cite{cresti2008valley,akhmerov2008theory}. To put these ideas into a realizable footing, we propose a multi-terminal p-n junction (PNJ) device structure and delve into the physics of complexities such as modal discrepancies at multiple terminals, related Dirac fermion optics, angle selective transmissions and edge state localizations on the valley filter operation.\\
\indent The proposed multi-terminal device, schematized in Fig.~\ref{fig:Fig.1}(a), exhibits the two independent operating regimes of valley polarization. In Regime 1, each terminal is designed to allow a specific valley component to transmit, depending on the edge state slope polarity, enabling each terminal to acts as a valley selective transmitter. The PSVF regime on the ohter hand is based on the parity of the number of zigzag atomic rows giving rise to the valley filter effect. The geometrical parameters of the device influence the polarization in this regime, which reveal intriguing physics that contribute to it.\\
\indent The paper is organized as follows. Section \ref{Sec2} describes the modeling and simulation of the device in detail. Section \ref{Sec3} is devoted to the results and discussion, with Sec. \ref{Sec3} A discussing in detail the polarization in Regime 1 and Sec. \ref{Sec3} B discussing the polarization in Regime 2. Section \ref{Sec3}B also provides insights into the optimization of the maximum polarization conditions. The effects of varying the input width with respect to the width of the scattering region and the variation of the lengths of the p and n regions are analyzed. Finally, the main conclusions and outlook are sketched out in Sec. \ref{Sec4}.\\
\section{Device and simulations details} \label{Sec2}
\begin{figure}[!htpb]
	\includegraphics[width=\linewidth]{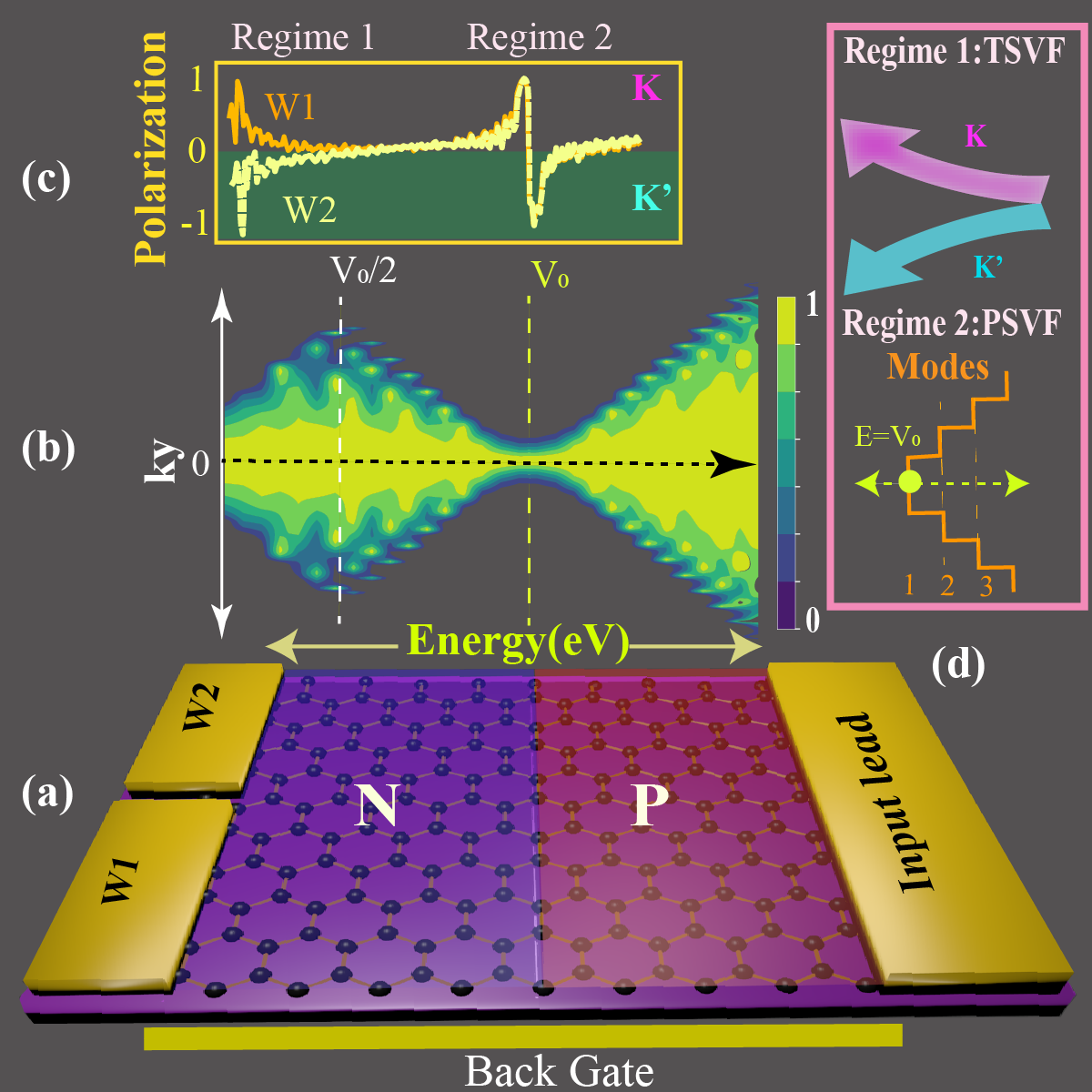}
	\caption{Overview (a) Schematic illustration of the proposed device comprising one input lead, two output leads, the device region with electrostatically doped p- and n- regions, and the back gate. The leads and the scattering region are anti-zigzag graphene while the output lead $W_2$ is bearded graphene. (b) Transmission spectrum as a function of $k_y$. The maximum transmission value is unity and is depicted by a bright yellow color, while reflection is shown by a deep blue color. (c) The device shows two regimes of polarization - Regime-1 and Regime-2. (d) Regime 1 named as terminal specific valley filter is due to band structure engineering (TSVF) and Regime 2: Parity specific valley filter (PSVF) is due to the odd parity of zigzag atomic rows along the width governed by the parity selection rules.}
	\label{fig:Fig.1}
\end{figure}

\indent The schematic diagram of the proposed device is shown in Fig. \ref{fig:Fig.1}(a). It consists of one input lead, two output leads ($W_1$ and $W_2$), and a scattering region. The scattering region, the input lead, and the output lead $W_1$ consist of anti-zigzag graphene, while the output lead $W_2$ consists of bearded graphene. The width and length of the scattering region are $70~ nm$ and $100 ~ nm$, respectively, with $N \sim 329$ (where $N$ is the number of zigzag rows of atoms). For odd values of $N$, this configuration is called the anti-zigzag configuration. The width of the input lead is generally equal to the width of the scattering region. The width of the two output leads is about half the width of the scattering region. The scattering region consists of an abrupt PNJ, which is electrostatically doped with Fermi level energy $E_F=0.25~eV$ on both the p and n sides. The back gate is used to tune the energy range for device operation.
The simulations are carried out using the Kwant software package \cite{groth2014kwant} based on scattering matrix formalism. The Hamiltonian represented in the second quantization form is constructed using the tight-binding model for graphene given by, 
\begin{equation}
\hat{H} = -t \sum_{i,j} \hat{c}_i^ {\dagger} \hat{c}_j + h.c. ,
\end{equation}
where, $t$ is a hopping integral across the nearest neighbour $i,j$ with the energy equal to $2.7~eV$.  Here, h.c., stands for the hermitian conjugate, $\hat{c}^{(\dagger)}_{i}$ is an annihilation (creation) operator of an electron on-site $i$ and the value of a lattice constant $a$ is equal to $0.246~nm$. \\
\indent The valley resolved transmission is calculated for both the output leads $W_1$ and $W_2$, which are annotated as $T_{K_{1}}$ and $T_{K^{\prime}_{1}}$ for $K$ and $K^{\prime}$ in $W_1$ and $T_{K_{2}}$ and $T_{K^{\prime}_{2}}$ in $W_2$, respectively \cite{jana2022robust}. The polarization $P_{1,2}$ in the leads $W_1$ and $W_2$ is given by
\begin{eqnarray}
 P_{1,2} =\frac{T_{K_{1,2}}-T_{K^{\prime}_{1,2}}} {T_{K_{1,2}} + T_{K^{\prime}_{1,2}}}.
\end{eqnarray}
The total transmission $T$ through the device is given by the sum of both $K$ and $K^{\prime}$ components from both outputs leads $W_1$ and $W_2$, given by
\begin{eqnarray}
   T= T_{K_{1}} + T_{K_{2}} + T_{K^{\prime}_{1}} +T_{K^{\prime}_{2}}.
\end{eqnarray}
Using this formulation, we compute the transmission across the output leads of the device for various parameters. Figure \ref{fig:Fig.1}(b) shows the transmission spectra for one input/output lead PNJ configuration. The device is translationally invariant along the $y$ direction, thus conserving the $k_y$ component of the wave vector while transmitting across PNJ. A carrier with normal incidence tunnels through the barrier irrespective of its height and width due to the pseudospin conservation. This peculiar transport results in zero backscattering for normal incidence, known as Klein tunneling \cite{allain2011klein,beenakker2008colloquium,katsnelson2006chiral}. From the transmission plot, it can be inferred that the spectra of $k_y$ values increase with the increasing energy in the low energy regime. When the incident electron's energy is equal to half the barrier height, which is $(V_0/2)$, the conditions for the evanescent wave and critical angle are met. Thereby, with further increase in energy, higher $k_y$ components in transmission are suppressed. At the energy equal to barrier energy$(V_0)$, also known as grazing energy, only normal components \emph{i.e.},  $k_y=0$ transmit with unity probability otherwise transmission is through evanescent waves. Around the grazing energy, the device shows polarization in Regime 2.\\
\indent Figure \ref{fig:Fig.1} (c) shows the polarization with a clear demarcation between two polarization regimes. The polarization in Regime 1 is due to the band structure engineering (TSVF), while that in Regime 2 is due to the odd parity of the zigzag ribbon governed by the parity selection rules (PSVF), as summarized in Fig.\ref{fig:Fig.1}(d).  The polarization at two output leads in Regime 1 is opposite, \emph{i.e.}, one corresponds to $K$ and the other to $K'$ due to output lead configuration. In Regime 2, both output leads correspond to the same polarization, which may be either $K$ or $K'$ depending on whether the energy level is above or below the Fermi level. The interesting transport physics in each of the polarization regimes will be discussed henceforth.
\section{Results and Discussion} \label{Sec3}
\subsection{Regime 1: Terminal specific valley filter(TSVF)}
The valley selective transmission at the output leads over a given energy range is attributed to the band structure engineering. To understand the transport physics, it is crucial to discuss the lattice structure and the corresponding band structure properties of the graphene nano-ribbons types used. As mentioned earlier, two types of lattice configurations are used in the proposed device: (1) an anti-zigzag and (2) bearded zigzag graphene. 
\begin{figure}[!htbp]
	\includegraphics[width=\linewidth]{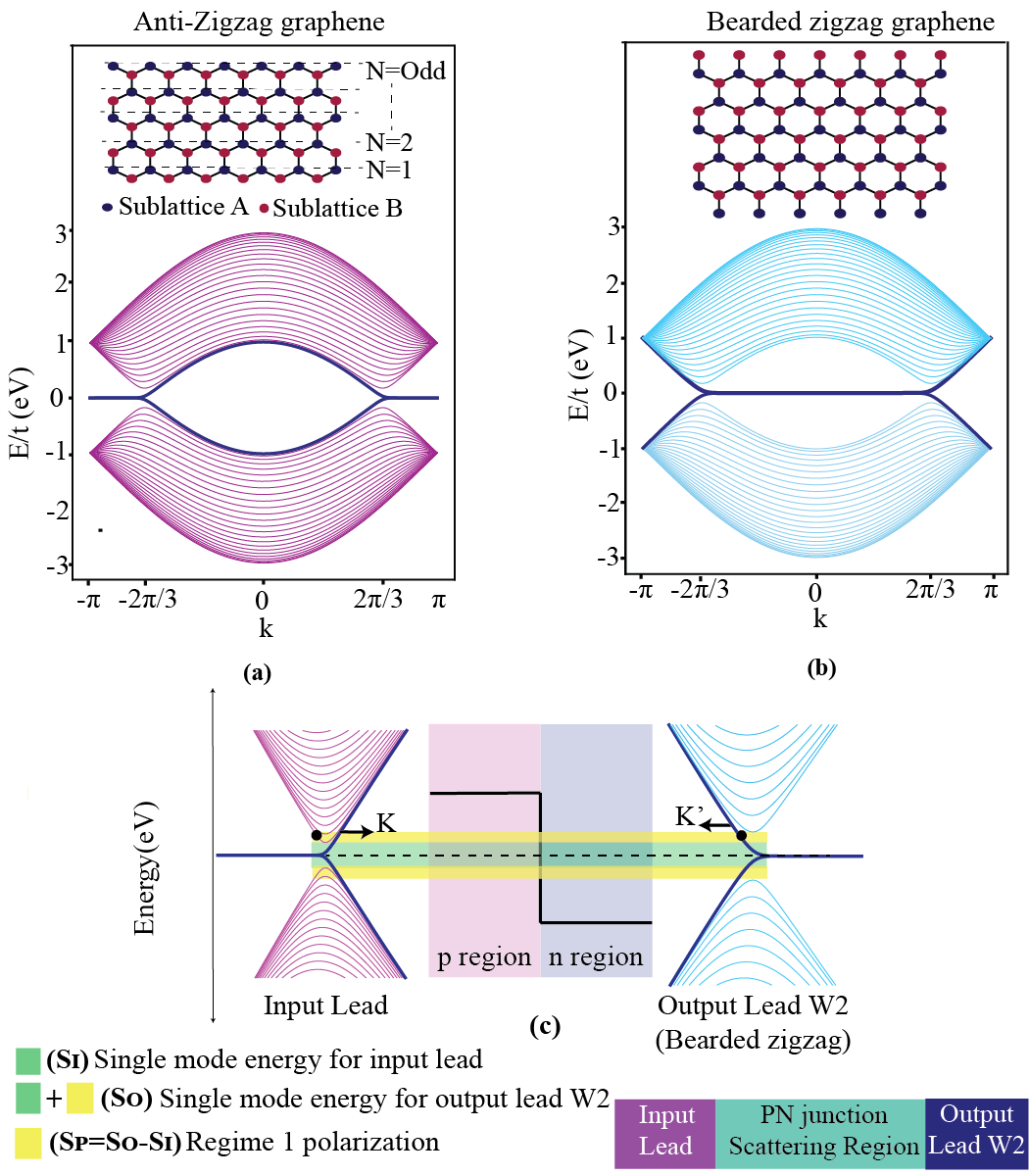}
	\caption{ The TSVF regime : Bandstructures of (a) zigzag and (b) bearded graphene with corresponding lattice configurations shown in the insets. The range of $k$ over which the flat band exists is depicted by dark blue lines and is different for each of the lattice types. (c) Schematic of the transport through the device for output lead $W_2$, made up of bearded graphene and half the width of the input lead. The green-shaded region shows the single-mode energy range for the input lead. The single mode energy range for the $W_2$ lead comprises the yellow and the green shaded areas. This range is larger than that for the input lead due to its smaller width.}
	\label{fig:Fig.2}
\end{figure}
\indent Figure \ref{fig:Fig.2}(a,b) shows the lattice structures and the computed $E-k$ relations for each of the configurations. The zigzag lattice with an odd number of zigzag atomic rows (N) is known as anti-zigzag graphene. The difference between a zigzag (N is even) and an anti-zigzag (N is odd) GNR lies in the transport properties of the edge states across the pn interface. The lattice structure is called a bearded zigzag lattice when the Klein nodes are attached to the zigzag-edged graphene. Both lattice structures have different sublattice atoms at the two edges, and the finite overlap of these edge wave functions gives rise to the specialized edge states. In Fig. \ref{fig:Fig.2}(a) and (b), the edge states are highlighted by dark blue lines, while the bulk states are depicted by the magenta and light blue color lines, respectively.\\ 
\indent These specialized edge states manifest as a partially flat region over a finite $k$ range, giving rise to a substantial density of states (DOS) at the Fermi level. The range of $k$ over which flat-band exists varies for each lattice type, as observed in the $E-k$ dispersion plots. For the zigzag edge GNR, edge states are completely flat over the $2\pi/3<\lvert{k}\rvert<\pi$ range and are dispersive for $k$ values in the first Brillouin zone. The range of $k$ over which edge states are completely flat for the bearded zigzag GNR is $0<\lvert{k}\rvert<2\pi/3$. The energy range over which only edge states conduct is between the bulk states, \emph{i.e.}, the second lowest of the conduction band and the second highest energy state of the valance band. Over this energy range, the edge states provide a single mode for conductance. The energy range is called a single mode energy (SME) range. The SME range is dependent on the physical dimension of the device, like width, and is given by, \cite{wakabayashi2001electronic}
\begin{eqnarray}              
\Delta_{S} ={4t\cos{\left( \frac{N-1}{2N+1}\pi\right)}}.
\end{eqnarray}
\indent We schematically present the transport mechanism when anti-zigzag (input) and bearded( output) graphene leads are adjoined at either side of a pn-junction in Fig. \ref{fig:Fig.2}(c). It shows the band structure at the input side (region 1 on the left), scattering region (region 2 in the middle), and band structure of an output lead (region 3 on the right). The SME range for an input lead $S_I$, shown by the green shaded energy range, is smaller as compared to the SME of output lead $ S_O$ shown by the yellow+green shaded area. The energy range shown by the green color in regions 1 and 3 corresponds to single-mode operation, whereas region 2 has multiple modes because of doping.\\
\indent Figure \ref{fig:Fig.3} shows the total transmission of the proposed structure and valley polarization. The total transmission as a function of the energy of the incoming electron is plotted in Fig. \ref{fig:Fig.3}(a). In the SME regime (green color), the total transmission through the device is unity combining both the output leads. While in the multi-mode energy (MME) regime (yellow color), the total transmission through the device is two as both the output leads having a single mode available contribute to a unity value to the total transmission. It is worth noting that the transmission value is dominated by the region with the lowest number of modes.
\begin{figure}[h!]
	\includegraphics[width=\linewidth]{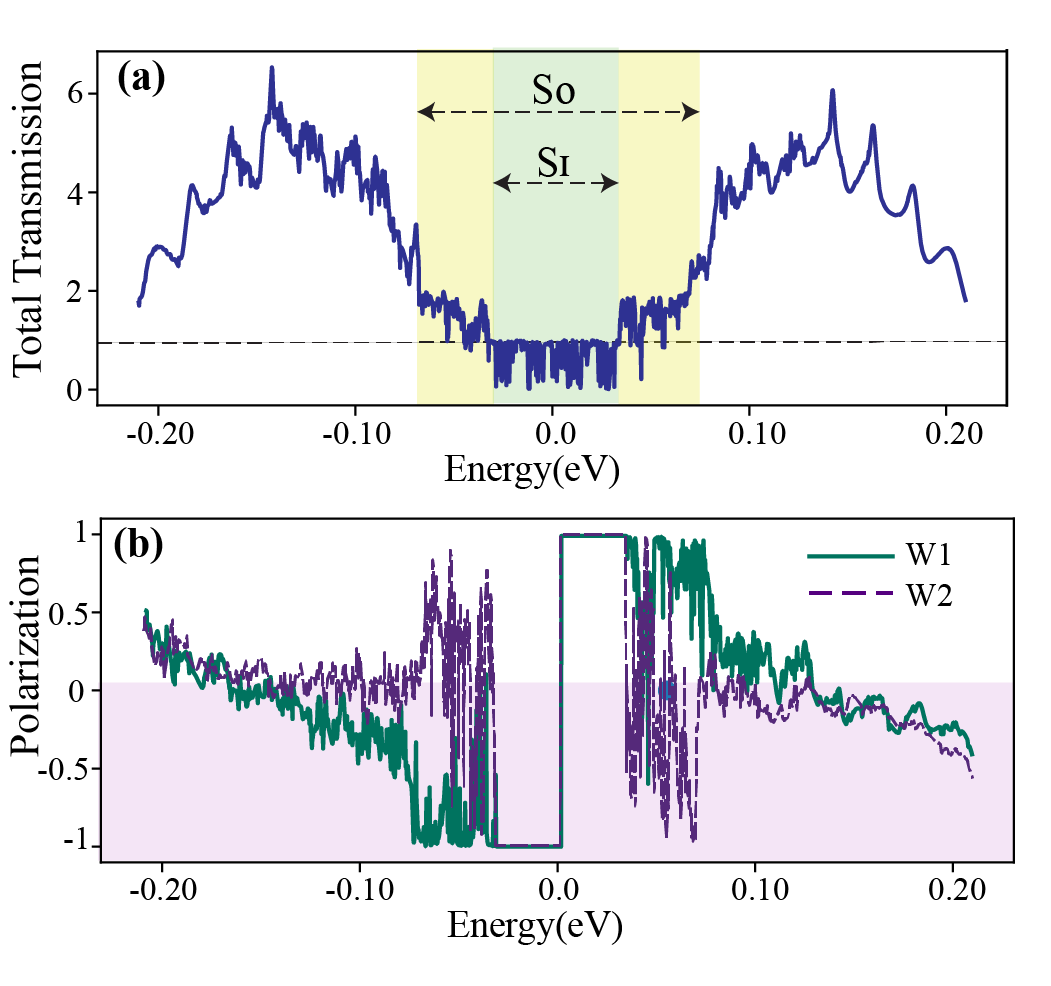}
	\caption{Transmission and valley polarization in the TSVF regime. (a) Total transmission plot and (b) polarization plot as a function of electron energy. The single mode energy range for input lead $S_I$ (green) and output lead $S_O$ (yellow+ green) are highlighted. In the $S_I$ range, the polarization is the same for both output leads. For the $S_O$ energy range, the polarization is different in each output lead.}
 \label{fig:Fig.3}
\end{figure}\\
\indent The dip in the total transmission plot is due to backscattering, which is prominent near zero energy. The backscattering occurs due to a mismatched boundary across the hetero-interface between zigzag and bearded graphene. The backscattering decreases the conductance and degrades the polarization value at the output lead $W_2$ \cite{wakabayashi2001electronic}. Figure \ref{fig:Fig.3}(b) shows the polarization plot that manifests the valley-selective functionality at the output leads.\\
\indent The terminal specific valley transmission depends on the polarity of the edge state slope in the $E-k$ dispersion of an output lead. For the output leads $W_1$ (anti-zigzag) and $W_2$ (bearded-zigzag), the polarity of an edge state slope at the Dirac points (\emph{i.e} at $\lvert{k}\rvert=2\pi/3$) are opposite as shown in \ref{fig:Fig.2}(a and b) by dark blue lines. In the energy range marked by the yellow color in Fig.\ref{fig:Fig.2}(c), the input lead operates in the multi-mode state with both positive and negative moving states. While the output leads are still in the SME range. A multi-mode operation is also observed across the PNJ due to significant doping. Therefore, a single allowed transmitting state at each output is either positive ($K$) or negative ($K'$) based on the  positive or negative edge state slope. Due to the opposite polarity of the edge state slope, two output leads can transmit two valleys in the yellow-marked energy range.
\subsection{Regime 2: Parity specific valley filter in multi-mode setup }
\begin{figure}[!htbp]
	\includegraphics[width=\linewidth]{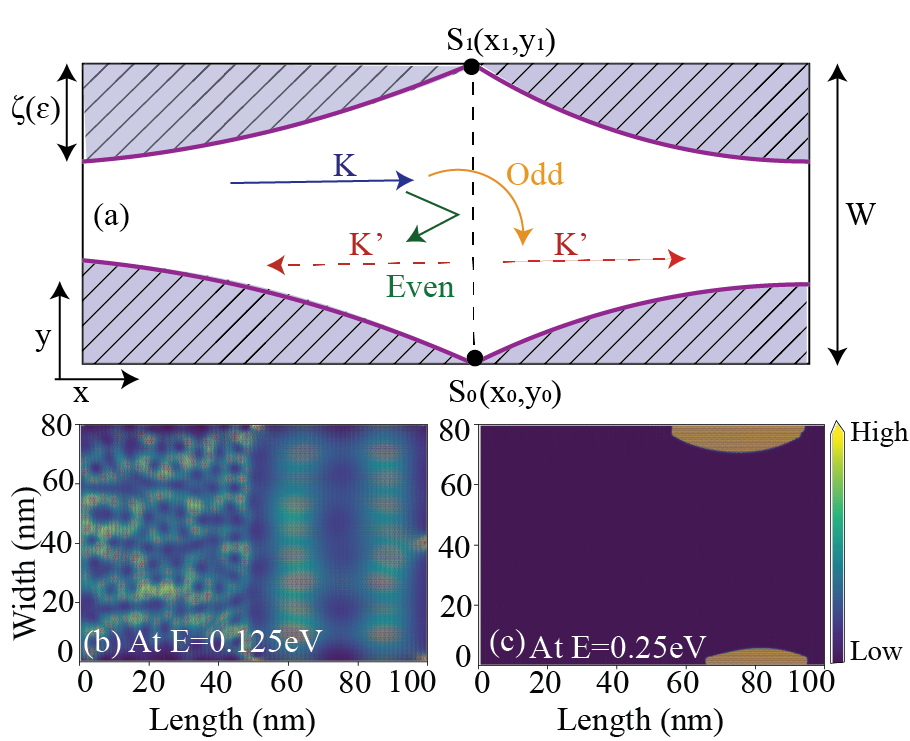}
	\caption{The PSVF regime: parity selective transmission condition and LDOS plots. (a) The shaded region shows the transverse extent $\zeta$ of the edge states in the device as a function of the device width $W$ and the electron kinetic energy, $\varepsilon$. At zero kinetic energy, states are localized at the edges, and the black dots show the span of wave vector interval. For zigzag graphene, with $N$ being even, the incoming $K$ state is reflected as a $K’$ state, giving rise to the valley valve effect. While for an odd value of $N$, the incoming state is transmitted as a $K’$ state, leading to the valley filter effect for the lowest energy mode. (b) LDOS plot at $E=0.125~eV$ and at (c) $E=0.25~eV$ shows different distributions; LDOS is accumulated at the device edges for $E=0.25~eV$.}
    \label{fig:Fig.4}
\end{figure}

\begin{figure}[!htbp]
	\includegraphics[width=\linewidth]{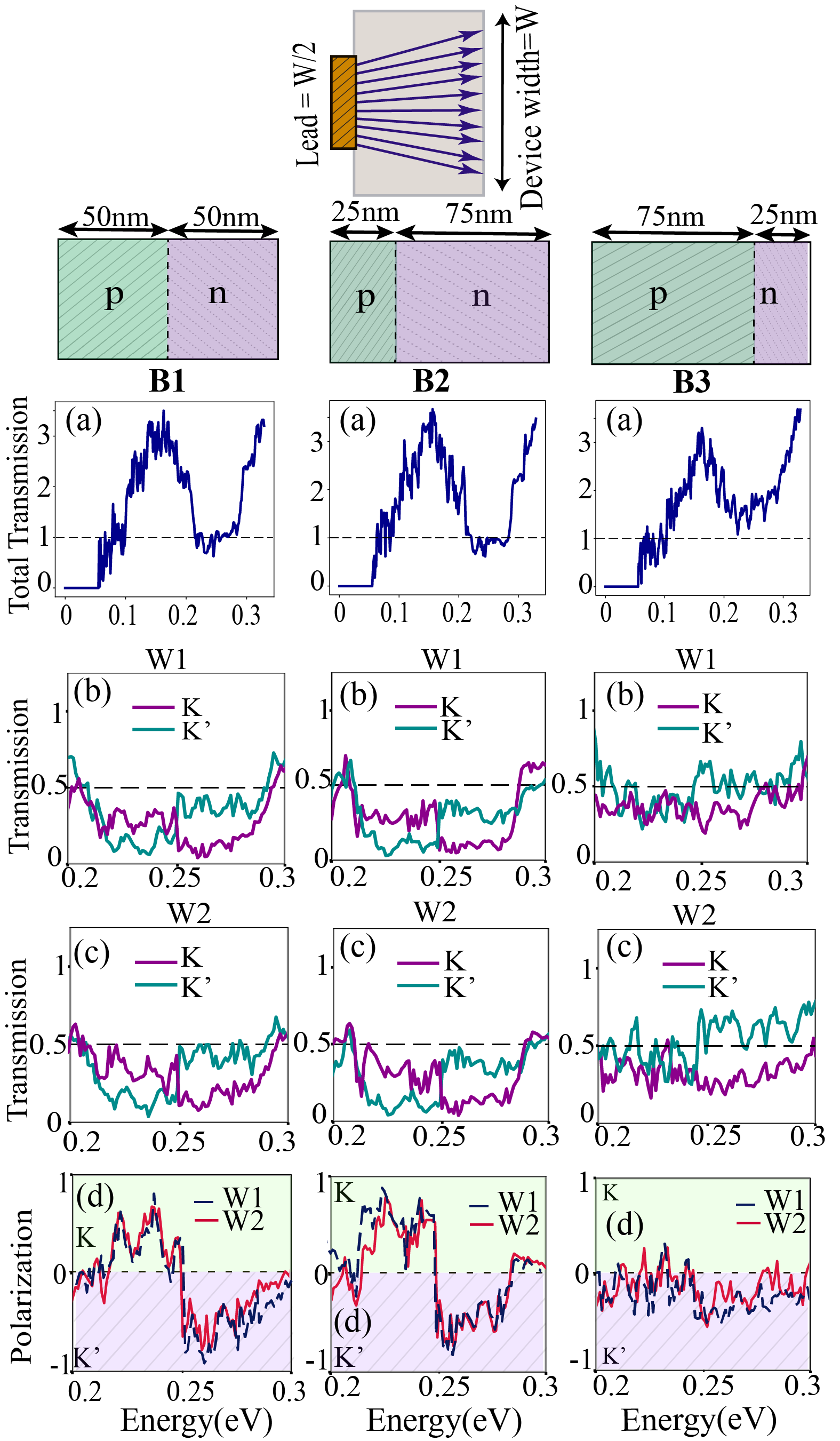}
\caption{Case A: Width of input lead equal to the width of the scattering region. The ray trace shows that the carrier trajectories coming from the input lead are parallel throughout the width of the device. With respect to input lead, three junction positions at A1: $50~nm$, A2: $25~nm$, and A3: $75~nm$ have been considered. For each of these device configurations, (a) total transmission plot, transmission for $K$ and $K'$ at output leads (b) $W_1$ and (c) $W_2$, and (d) polarization are shown. For the positive energy range, the length of n region affects the transmission and polarization value. The length of n region is sufficiently large for devices A1 and A2. Over the single-mode energy range, the total transmission value is one and attains the maximum polarization. In device A3, the length of the n-region is small and due to the quantum tunneling, the total transmission value is  above one, and the overall polarization value degrades.}
    \label{fig:Fig.5}
\end{figure}

\begin{figure}[!htbp]
	\includegraphics[width=\linewidth]{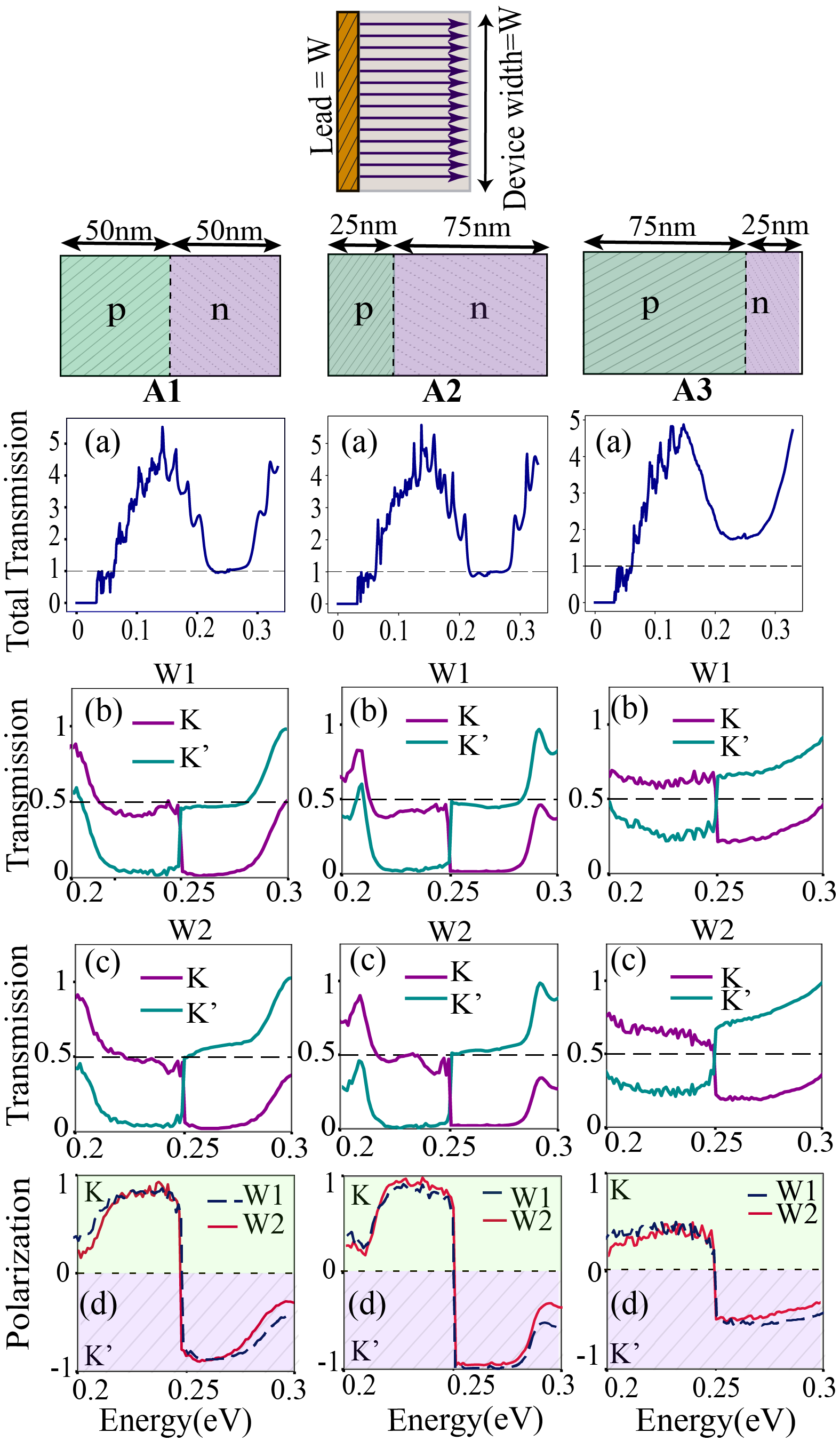}
	\caption{Case B: The width of input lead is half the width of the scattering region. The ray trace shows that the carrier trajectories coming from the input lead spread out across the width of the device. Similar to the earlier case, three device configurations with junction positions at B1: $50~nm$, B2: $25~nm$, and B3: $75~nm$ have been considered. For each of these device configurations, (a) total transmission plot, transmission for $K$ and $K'$ at output leads (b) $W_1)$ and (c) $W_2$, and (d) polarization are shown. The total transmission value is lower than in the previous case due to the lesser number of modes present at the input lead. For devices B1 and B2 (wider n-region), the total transmission value is approximately one in the single mode energy range while for device B3 (narrower n-region) it is above one due to the quantum tunneling. There is a finite transmission for $K$ and $K’$ in each case, and so the polarization degrades. The worst-case polarization is observed for device B3 due to the combined effect of dispersed carrier trajectories and quantum tunneling.}
 \label{fig:Fig.6}
\end{figure}

Inter-Valley scattering is typical to graphene PNJs with zigzag edges, regardless of whether the interface region is sharp or atomically smooth \cite{akhmerov2008theory}. The reason for inter-valley scattering is localized edge states, which are the lowest energy modes. The localized edge states have incoming and outgoing states originating from different valleys. The scattering over PNJ must change valleys in order to maintain the current flow. In our simulations, we consider a sharp PNJ profile. The edge states are completely localized at the zigzag edges of the device. The transverse extent $\zeta$ of the edge states along the width of the device depends on the kinetic energy of the incoming electron $\varepsilon$, which is given by
\begin{eqnarray}
\varepsilon= E_F-V_0
\end{eqnarray}
\begin{eqnarray}
\zeta(\varepsilon)= W/\ln{\lvert{\varepsilon W/ v_f \hbar\rvert}}
\label {trans}
\end{eqnarray}
\indent The shaded region with hashed lines in Fig. \ref{fig:Fig.4}(a) shows the transverse extent $\zeta$ of the edge state wave vector $k$ as a function of kinetic energy $\varepsilon$ and the device width $W$ as given in \eqref{trans}. Here, $E_f$ is the Fermi energy, $V_0$ is the barrier height, $\hbar$ is Planck’s constant, and $v_f$ is the Fermi velocity. The width $\zeta$ decreases as the kinetic energy decreases and attains its minimum at zero kinetic energy. This minimum is of the order equivalent to the lattice constant $a$ as shown by the points $S_0$ and $S_1$ in Fig.\ref{fig:Fig.4}(a). The wave vector of the edge state spans the interval of order $1/a$ between $K$ and $K’$, thereby facilitating inter-valley scattering processes. An incoming state, say, $K$ (blue) at the interface (dashed line), switches to $K’$ state while being transmitted or reflected, assisted by the inter-valley scattering.
\noindent The transmission or reflection of the incoming state depends on the parity of $N$. For even $N$, the parity of sub-lattice states across the PNJ is not the same, thus, the inter-valley scattering reflects an incoming state. This configuration manifests the valley valve effect for the lowest energy state. While in the case of odd $N$, the sub-lattice states across PNJ are of the  same parity, giving rise to the valley filter effect by transmitting an incoming state. The complete derivation and discussion regarding the parity selection rule are elaborated in \cite{cresti2008valley,akhmerov2008theory}.\\
\indent The LDOS plots at two different carrier energies are shown in Fig. \ref{fig:Fig.4}(b-c). The LDOS is uniformly distributed in the device at $E = 0.125~ eV$ as seen in Fig. \ref{fig:Fig.4}(b), whereas at the grazing energy of $E = 0.25 ~eV$, the charge density is accumulated at the edges of the device, as depicted in Fig. \ref{fig:Fig.4}(c). It can be inferred that the main contribution to the conductance is due to the edges for a single-channel energy range. Thus, interesting polarization effects or even complete valley-switching of the electronic states can be realized by tailoring the edge states at the input and output leads of a graphene PNJ. We hence focus our discussion on how the width of the input lead and the lengths of p- and n-doped regions influence the transmission and valley polarization in the regime of multi-mode operation. The multi-mode state is referred for the input side. %In earlier works \cite{rycerz2008nonequilibrium,akhmerov2008theory,rycerz2007valley}, a device with a physical constriction and single operating mode was considered to study the valley valve/filter effect. In this work, we consider a nanoscale device operating in a multi-mode state. 
Although at grazing energy, the device operates with a single mode, the degree of polarization has a strong dependence on the way input modes approach the PNJ. So, it is essential to consider the behavior of carriers in graphene for analyzing the way input modes approach the interface.\\
\indent Low-energy carriers in graphene show Dirac fermion behavior, and therefore, in a PNJ exhibit electronic analogs of Snell’s law and Veselago lensing. Following the optics analogy, the input lead in the multi-mode setup acts as an optical slit, and its width determines the angular dispersion of the incoming modes at the PN interface. We consider two cases - Case A: When the width of the input lead is equal to the width of the device as illustrated in Fig. \ref{fig:Fig.5} and Case B: When the input lead width is half and centered along the device width as seen in Fig.\ref{fig:Fig.6}. The ray trace in Fig.\ref{fig:Fig.5} shows that all the electron trajectories are parallel to the device edges and are perpendicular to the PNJ interface throughout the width of the device. While the ray trace in Fig.\ref{fig:Fig.6} shows the ray dispersion along the lead edges. The rays are perpendicular in the middle region and diverge out towards the edges along the PN interface (see Appendix A).\\
\indent The degree of valley filtering also depends on the doping profile in the PNJ. Firstly, the doping level decides the energy range where polarization is observed. Secondly, the length of doped regions is equivalent to the potential barrier width in respective energy ranges. The narrow barrier width allows conventional tunneling through it. Thus, here we have considered the effect of barrier width variation along with the input lead variation for examining the conditions to achieve the maximum polarization.\\
\indent For the doped region length variation, the transmission in the device simplifies to the transport through the potential barrier in graphene. In the positive energy range, the length of n- region acts as the barrier width and for the negative energy range, the p- region length acts as the barrier width. In a graphene PNJ, the transmission probability of an incoming wave depends on the angle of incidence with respect to the PN interface. Above a certain angle, the incident wave is totally reflected at the critical angle. The critical angle condition is formed when the energy of an incoming electron $E$ is greater than half the barrier height. When the critical angle condition is met, the evanescent waves are formed and total internal reflection results in total reflection for the PNJ. But for the potential barrier, evanescent waves appear in the barrier region and give rise to quantum tunneling \cite{allain2011klein}. So, the narrower the barrier region, higher the quantum tunneling, which comprises the tunneling of both the $K$ and $K’$ components. This has direct implications on the degree of polarization at the output lead.\\
\indent We now elaborate on the results shown in Fig. \ref{fig:Fig.5} and Fig. \ref{fig:Fig.6}. The position of the PNJ is varied along the device length. We have considered three device configurations represented as A(B)1, A(B)2, and A(B)3, wherein the junction is positioned at $50~nm$, $25~nm$, and $75~nm$, respectively, from input lead. The total transmission plots, \emph{i.e.}, the sum of $K$ and $K'$ transmission components from both output leads, are shown by A(B)i(a), where i=1,2,3. The A(B)i(b and c) show the transmission of $K$ and $K'$ components in output lead $W_1$ and $W_2$, respectively. And finally, the A(B)i(d) shows the polarization at both output leads.\\
\indent The total transmission plots for all three cases show an increase in transmission value with an increase in electron energy as a number of modes get added. When the electron energy reaches half the potential height. \emph{i.e.}, $0.125~eV$, the critical angle condition is formed and the higher angular modes are suppressed. When the energy of the incoming carrier approaches the potential barrier height \emph{i.e.}, $0.25~eV$, only edge states conduct and pseudo-spin conservation is valid only for the normal incident case. Thus, in SME the total transmission value for A1(a) and A2(a) is unity, while for case A3(a), it is higher than one and the transmission curve is smooth which shows that the transmission is via quantum tunneling due to a narrower n- region. With a pseudo-spin conserved transmission, the total transmission in the SME range is one. For two output leads, the maximum transmission value is half as observed in A1(b and c) and A2(b and c). But for A3(b and c) the  value is slightly higher for both $K$ and $K'$ components, which in turn degrades the polarization. The polarization is maximum for A2(d) and slightly lower for A1(d). The polarization dropped to nearly half its maximum value for case A3(d) due to the quantum tunneling.\\
\indent For case B, \emph{i.e.}, when the input lead width is half the scattering region width, the results are similar to that in the previous case. But, now as the input lead width is half that in the previous case, the maximum total transmission is lower. This is because of less number of modes present in the input lead. As of the previous case, the transmission in cases B1(a) and B2(a) in the single mode energy range is approximately equal to one while in B3(a) is higher because of quantum tunneling due to the narrower n region. The transmission of $K$ and $K’$ in $W_1$ and $W_2$, plotted in B1(b and c) and B2(b and c) shows the flickering curve due to the quantum tunneling because of reduced wave function amplitude at edges and angle selective transmission at PN interface (Appendix A). In quantum tunneling, there is a finite transmission probability of both valleys and thus, the polarization in B1(d) and B2(d) is lower as compared to A1(d) and A2(d) cases. The worst polarization B3(d) is due to the combined effects of narrower n region and dispersed electron trajectories due to input lead width variation.\\
\indent Thus, the conditions for achieving maximum polarization could be summarized as (1) Zigzag-edged graphene should have odd parity for $N$ to allow inter-valley scattering across the PNJ interface and achieve polarization at the lowest energy mode (2) The edge states conducting during a single mode energy range are completely localized at the physical device edges \cite{nakada1996edge,fujita1996peculiar}, and (3) At grazing energy, only normally incident electrons have complete transmission probability, otherwise, the transmission is via evanescent waves or through quantum tunneling. So, to get maximum polarization, the electron should be incident normally at the PN interface and especially at the edges of the device where LDOS is concentrated. If an electron is an incident at an angle other than the normal angle, firstly because of the diffraction effect the amplitude is reduced, and secondly, due to angle selective transmission across the PN interface at grazing energy, the quantum tunneling probability increases.\\
\section{Conclusion} \label{Sec4}
\indent In this work, we proposed an all-electrical valley polarizer using zigzag edge graphene nanoribbons in a multi-terminal device geometry, that can be gate-tuned to operate along two independent regimes: (i) a terminal-specific valley filter that utilizes bandstructure engineering, and (ii) a parity-specific valley filter that exploits the parity selection rule in zigzag edge graphene. We showed that the device exhibits intriguing physics in the multimode regime of operation that affects the valley polarization and hence investigate various factors affecting the polarization in wide device geometries, such as, optical analogs of graphene Dirac fermions, angle-selective transmission via p-n junctions, and the localization of edge states. We optimized the geometry of the proposed device to achieve maximum valley polarization, thereby, paving the way toward a physics based tunable valleytronic device design using monolayer graphene. The proposed concepts can also be extended to the photonic crystals \cite{gao2017valley}  for the experimental implementation of a valley polarizer without the requisite of polarized light.  %In Regime 2, the parity specific valley filtering was observed. The odd parity of a number of zigzag atomic rows contributed to the valley filtering effect. In earlier related works, the idea was implemented with single-moded device dimensions; here, we implemented it on a large-dimension device. The simulation results showed a significant effect on the degree of polarization due to the physical device parameters like input lead width and the doped region length. 
\begin{acknowledgments}
The authors acknowledge the Science and Engineering Research Board (SERB), Government of India, for Grant No. STR/2019/000030 and Grant No. CRG/2021/003102, and the Ministry of Education (MoE), Government of India, Grant No.STARS/APR2019/NS/226/FS under the STARS scheme
\end{acknowledgments}

\appendix

\section{Diffraction}
\begin{figure}[!htbp]
	\includegraphics[width=\linewidth]{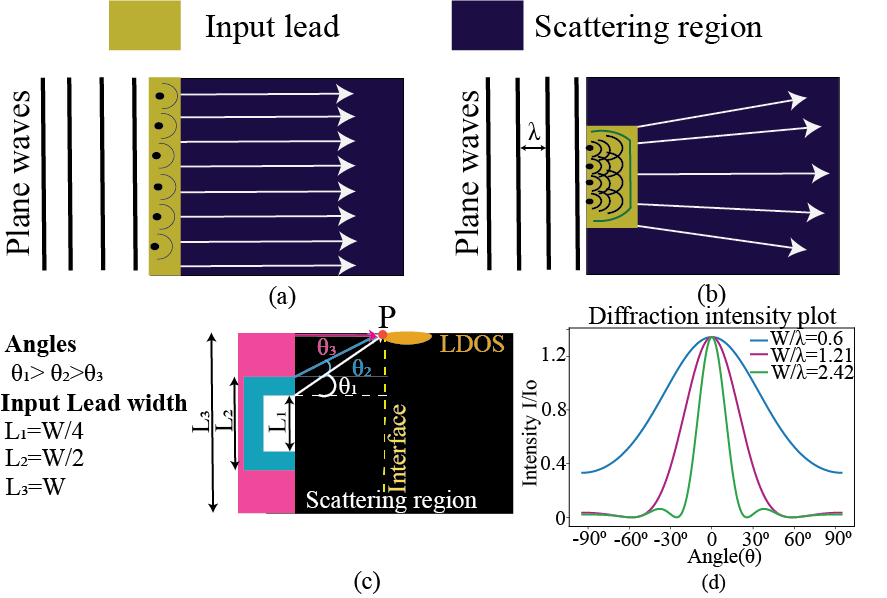}
	\caption{ Optical slit analogy a) No diffraction when input lead and scattering region have the same width. b) Diffracted ray pattern when input lead width is smaller than scattering region. c) The angle between the axis normal to interface from the lead top end point and the point P at interface endpoints at device edges decreases for increasing lead widths. d) Intensity distribution plot for different input lead widths at a constant wavelength.}
 \label{fig:Fig.7}
\end{figure}

Dirac fermions in graphene obey the laws of ray optics, such as, the Snell’s law and exhibit Veselago lensing. It is important to note that the incoming electron trajectory should have a finite angle with respect to the axis perpendicular to the PN interface for exhibiting the Veselago lensing and negative refraction effect. However, for the normal incidence, fermions just transmit through the barrier following Klein tunneling. The angular dispersive behavior of graphene fermions is the result of diffraction \cite{darancet2009coherent,lee2015observation}. \\
\indent In our work, we show that the input lead acts as an optical slit for incoming modes in the scattering region. When the width of the input lead is equal to the scattering region width, there is no diffraction and all modes are parallel to each other Fig.\ref{fig:Fig.7}(a). While in the case of the smaller width of the input lead the electron trajectories show a diffraction effect. There is bending at the edges of the input lead for an incoming mode and results in angular dispersion of the electron trajectory Fig.\ref{fig:Fig.7}(b).\\
\\
\indent For an incident intensity of $I_0$, the intensity distribution in the diffracted orders is given by \cite{young2015university},
\begin{eqnarray}              
 \ I= I_0 \left[ \frac{sin(\beta /2}{\beta/2} \right ]^2
\end{eqnarray}
where, $W$ is the width of the slit, the wavelength of the Dirac fermion with energy $E$ is $\lambda=h\nu_f/E$, where, $\nu_f=10^6 m/s$, and the angle $\theta$ is the angle from the point of interest $P$ (in red)\ref{fig:Fig.7}(a) to the axis perpendicular to the slit plane shown by the white dashed line. The $\beta$ is the total phase difference between the two rays approaching from the top and bottom of the slit to point $P$.
The path difference between the ray from the top of the slit to the ray in the middle of the slit with width $W$ is $ W/2sin\theta$. Thus, $Wsin\theta$ is equal to twice the difference between the top and bottom rays. The phase difference is $2\pi/\lambda$ times the path difference. Thus, the total phase difference can be written as $\beta=2\pi/\lambda*(W sin\theta)$.
The expression of intensity in terms of angle $\theta$ is, 
\begin{eqnarray}              
 \ I= I_0 \left[ \frac{sin[\pi W(sin\theta)/\lambda}{\pi W(sin\theta)/\lambda} \right ]^2
\end{eqnarray}
The plot \ref{fig:Fig.7}(d) shows the intensity distribution as a function of angle $\theta$ and the $W/\lambda$ ratio. Here, we considered the value of $E=0.25eV$, $\lambda$=16.52 nm, and varied the value of width $W$. The ratios $W/\lambda$=0.6, 1.21, 2.42 are plotted with pink, blue, and green respectively in \ref{fig:Fig.7}(d). From the plot, it can be observed that the curve for $W/\lambda$ =0.6 is widespread over the angular spectrum. As the width $W$ increases, the intensity gets maximized in the zeroth order suppressing higher-order side lobes.\\
\indent Consider an example with different widths of input lead L1, L2, and L3 with respect to scattering region width $W$ as shown in \ref{fig:Fig.7}(c). In Fig.\ref{fig:Fig.7}(c), the smallest slit gives the largest diffraction angle ($/theta_1$) with some amount of intensity being distributed in the corresponding diffracted orders (blue curve in \ref{fig:Fig.7}(d)). Thus, the electron wave function reaches point $P$ with a finite angle and reduced amplitude. Again by angular selective transmission at the PN interface, there is a further reduction in the transmission. At grazing energy, the conductance is mostly due to quantum tunneling. 

Similarly, if we consider the case with equal widths of both regions, there is no diffraction. At grazing energy due to anti-zigzag configuration, Klein tunneling contributes to the single mode conductance with a lesser probability of quantum tunneling.  Fig. \ref{fig:Fig.7}(a) and (b) show the ray trace for equal and smaller input lead width respectively, which  forms the basis for the explanation of the results.

\section{Transmission across PN interface under different scenarios}
 \subsection{Doped region length variation: Transmission across potential barrier}
The PNJ has highly angle-dependent transmission.  The transmission coefficient vanishes after a certain angle, called a critical angle. The condition for a critical angle is, 
\begin{eqnarray}              
 \sin\theta = \frac{V_0-E}{E}
\end{eqnarray}
beyond the critical angle, evanescent waves are present on another side of the junction and total reflection is observed. In the case of the potential barrier, once the condition of the critical angle is reached, the evanescent is formed on the other side of the junction, unlike the PNJ, in the barrier, there is no total reflection.  The presence of a second junction in the barrier case, allows the wave to be transmitted through the barrier via quantum tunneling with diminished amplitude. 
 \subsection{Quantum tunneling at grazing energy:}
The carrier arrives with energy $E$ which is exactly equal to barrier energy. The entire transmission of this energy is via evanescent waves except for normal incidence. The transmission probability decreases with an increase in incidence angle. Thus, for the maximum polarization, there must be a normal incidence in which pseudospin is conserved, while in quantum tunneling the transmission probability is the same for both $K$ and $K’$. Thus, the polarization is lowered. For a detailed explanation for transmission across PNJ and barrier refer\cite{allain2011klein}
Note: The polarization results for Regime 2 is valid with input and one output lead device setup also, while the polarization in Regime 1 requires a different output lead configuration for showing opposite valley polarization. 
% whether or not they are actually referenced in the text. This is an appropriate
% for the sample file to show the different styles of references, but the author's

\nocite{*}
\bibliography{apssamp}% Produces the bibliography via BibTeX.

\end{document}